\documentclass[12pt]{article}

\begin{document}
\begin{center}
{\bf Effective Lagrangian at Cubic Order in Electromagnetic Fields and Vacuum Birefringence}\\
\vspace{5mm}
 S. I. Kruglov \\
\vspace{5mm}
\textit{University of Toronto at Scarborough,\\ Physical and Environmental Sciences Department, \\
1265 Military Trail, Toronto, Ontario, Canada M1C 1A4}
\end{center}

\begin{abstract}
The effective Lagrangian of electromagnetic fields at the cubic
order in field strength is considered. This generalized Lagrangian
is motivated by electrodynamics on non-commutative spaces. We find
the canonical and symmetrical energy-momentum tensors and show that
the vacuum in the model behaves like an anisotropic medium. The
propagation of a linearly polarized laser beam in the external
transverse magnetic field is investigated. We obtain the dispersion
relation and refraction indexes for two polarizations. From
experimental values of the induced ellipticity, reported by PVLAS
collaboration, the constraint on parameters in the effective
Lagrangian is evaluated.
\end{abstract}

\section {Introduction}

The magnitude of the rotation \cite{Zavattini} and ellipticity
\cite{Zavattini1} (reported by the PVLAS Collaboration) of a
linearly polarized laser beam propagating through a transverse
magnetic field can not be explained within quantum electrodynamics
(QED) \cite{Adler}, \cite{Biswas}. This stimulates activities in the
theoretical proposals on physics beyond the Standard Model. Possible
explanations of the phenomena observed, in the framework of particle
physics, are in \cite{Sikivie}, \cite{Maiani}, \cite{Raffelt},
\cite{Gies}, \cite{Masso}, \cite{Brax}, \cite{Kruglov},
\cite{Beswick}. The effect of vacuum birefringence takes place also
in Lorentz violating electrodynamics \cite{Kostelecky}. Two popular
scenarios, explaining data of the PVLAS experiment, include the
existence of a new axion-like (spin-0) particle (ALP) \cite{Maiani}
and/or minicharged particles (MCPs) \cite{Gies} (see \cite{Ahlers},
\cite{Ringwald} for last reviews). The best fit is obtained for MCPs
of spin-1/2 \cite{Ahlers}. In addition, the parameters of ALP are
different as compared with the parameters of a QCD axion. The case
when the coupling and the mass of an axion-like particle depend on
the temperature and matter density was considered in \cite{Jaeckel}.
This can adjust astrophysical bounds to allow for the PVLAS signal.

In this letter, we study the induced ellipticity of the laser beam
(vacuum birefringence) using the effective Lagrangian at cubic order
in the electromagnetic field strength which is motivated by
electrodynamics on non-commutative (NC) spaces.

Models on NC spaces attract a great interest because NC coordinates
appear in the superstring theory with the presence of the external
background magnetic field \cite{Seiberg}. In the NC field theories
the Lorentz invariance is broken due to the fact that the constant
parameters $\theta_{\mu\nu}$ are coupled to tensors, but a twisted
form of the Lorentz invariance is valid \cite{Chaichian}. The NC
parameter $\theta$ is extremely small and astro-physical bounds on
$\theta^{-1/2}$ are of the order of the Planck scale. Experimental
bounds on the NC parameters are discussed in \cite{Carroll}.
Coefficients for the Lorentz violation in electrodynamics within the
Standard Model Extension (SME) are estimated from Cosmic Microwave
Background (CMB) radiation in \cite{Kostelecky1}. The Lorentz
violation constraints in the photon sector from measurements of the
linear polarization in gamma-ray bursts were considered in
\cite{Kostelecky2}.

We use the Heaviside-Lorentz system of units, and $\hbar=c=1$.

\section{Lagrangian and Field equations}

It was shown that there is no polarization rotation and ellipticity
observed in the NC version of electrodynamics \cite{Guralnic},
\cite{Cai}, \cite{Abe}, \cite{Zahn}, \cite{Mariz}, \cite{Chatillon}.
The effective Lagrangian of the NC version of electrodynamics, at
first order in the NC parameter $\theta$, is cubic in the
electromagnetic field strength. Therefore, to explain the data of
the PVLAS experiment, we consider the generalized effective
Lagrangian of electromagnetic fields at cubic order in the field
strength. Possible structures, including second-rank
``tensor"parameters in the Lagrangian, are as follows:
\[
\theta^{(1)}_{\alpha\beta}F_{\alpha\beta}F^2_{\mu \nu},~~
\theta^{(2)}_{\alpha\beta} F_{\mu \alpha}F_{\nu \beta}F_{\mu \nu},~~
\theta^{(3)}_{\alpha\beta}\widetilde{F}_{\alpha\beta}F^2_{\mu\nu},~~
\theta^{(4)}_{\alpha\beta}\widetilde{F}_{\alpha\beta}F_{\mu\nu}\widetilde{F}_{\mu\nu},~~
\]
\[
\theta^{(5)}_{\alpha\beta}F_{\alpha\beta}F_{\mu\nu}\widetilde{F}_{\mu\nu},~~
\theta^{(6)}_{\alpha\beta} \widetilde{F}_{\mu \alpha}F_{\nu
\beta}F_{\mu \nu},~~ \theta^{(7)}_{\alpha\beta} \widetilde{F}_{\mu
\alpha}\widetilde{F}_{\nu \beta}F_{\mu \nu},~~
\theta^{(8)}_{\alpha\beta} \widetilde{F}_{\mu
\alpha}\widetilde{F}_{\nu \beta}\widetilde{F}_{\mu \nu},
\]
where $F_{\mu \nu}=\partial_\mu A_\nu- \partial_\nu A_\mu $ is the
strength of the electromagnetic field,
$\widetilde{F}_{\mu\nu}=(1/2)\varepsilon _{\mu \nu \alpha \beta }F
_{\alpha \beta}$ $( \varepsilon _{1234}=-i)$ is a dual tensor. It is
easy to verify that these structures are not independent and may be
converted into only the first two terms. So, we have two independent
antisymmetric ``tensor"-parameters, or four three-``vectors". As
parameters $\theta_{\mu\nu}$ are not transformed as real tensors,
the Lorentz invariance is broken. Here we investigate the ``minimal"
extension of NC electrodynamics, and therefore, the possible
independent structure
\[
\theta_{\alpha\beta\mu\nu}F_{\alpha\beta}F_{\mu\gamma}F_{\nu\gamma}
\]
(including forth-rank ``tensor"-parameters $
\theta_{\alpha\beta\mu\nu}$) is not considered here.

The generalized effective Lagrangian of electromagnetic fields at
cubic order in the field strength is given by
\begin{equation}
 {\cal L}=-\frac14F^2_{\mu \nu}+\frac18\theta^{(1)}_{\alpha\beta}
 F_{\alpha\beta}F^2_{\mu \nu}-\frac12\theta^{(2)}_{\alpha\beta}
 F_{\mu \alpha}F_{\nu \beta}F_{\mu \nu}.
\label{1}
\end{equation}
Two constant ``tensors" $\theta^{(1)}_{\alpha\beta}$ and
$\theta^{(2)}_{\alpha\beta}$ are independent. At
$\theta^{(1)}_{\alpha\beta}=\theta^{(2)}_{\alpha\beta}$, one comes
to NC electrodynamics \cite{Bichl} (using Seiberg-Witten map
\cite{Seiberg})with the accuracy of ${\cal O}(\theta^2)$. The
Lagrangian (1) can also be rewritten as
\begin{equation}
 {\cal L}=\frac12 \left( {\bf E}^2-{\bf B}^2 \right)\left[1+({\bf
 \alpha}\cdot{\bf B})- ({\bf \xi}\cdot{\bf E})\right] -
\left({\bf E}\cdot{\bf B}\right) \left[\left({\bf \beta}\cdot{\bf
E}\right)+ ({\bf \gamma}\cdot{\bf B})\right]
 ,
\label{2}
\end{equation}
where the electric field is $E_i =iF_{i4}$ and the magnetic
induction field being $B_{i} =\epsilon_{ijk}F_{jk}$, ${\bf
\alpha}_i= 2{\bf \theta}_i^{(2)}-{\bf \theta}_i^{(1)}$, ${\bf
\beta}_i= {\bf \theta}_i^{(2)}$,
$\theta_i^{(1,2)}=(1/2)\epsilon_{ijk}\theta_{jk}^{(1,2)}$, and
$\gamma_i=i\theta_{i4}^{(2)} $,
$\xi_i=2i\theta_{i4}^{(2)}-i\theta_{i4}^{(1)}$. The parameters
$\theta^{(1,2)}_{\mu\nu}$ have the dimension of $(\mbox{length})^2$.
It follows from Eq.(2) that terms containing parameters ${\bf
\alpha}_i$,  ${\bf \beta}_i$ violate CP - symmetry. The
Lagrange-Euler equations lead to equations of motion as follows:
\[
\partial_\mu F_{\nu\mu}+\frac12\theta_{\alpha\beta}^{(1)}\partial_\mu
\left( F_{\mu \nu}F_{\alpha\beta}\right)+\frac14\theta_{\mu
\nu}^{(1)}
\partial_\mu\left(F^2_{\alpha\beta}\right)
\]
\vspace{-8mm}
\begin{equation}
\label{3}
\end{equation}
\vspace{-8mm}
\[
-\theta_{\nu\beta}^{(2)}\partial_\mu\left(F_{\alpha
\beta}F_{\mu\alpha}\right)+\theta_{\mu\beta}^{(2)}\partial_\mu\left(F_{\alpha
\beta}F_{\nu\alpha}\right)-\theta_{\alpha\beta}^{(2)}\partial_\mu\left(
F_{\mu\alpha}F_{\nu\beta}\right)=0 .
\]
Eq.(3) can be represented as
\begin{equation}
\frac{\partial}{\partial t}{\bf D}-{\bf \nabla}\times {\bf
H}=0,\hspace{0.3in} {\bf \nabla}\cdot{\bf D}=0 , \label{4}
\end{equation}
where the displacement (${\bf D}$) and magnetic (${\bf H}$) fields
are given by
\[
 {\bf D}=\frac{\partial {\cal L}}{\partial \textbf{E}}=\left[1+({\bf \alpha}\cdot
 {\bf B})- ({\bf \xi}\cdot{\bf E})\right]{\bf E}-\left[({\bf \beta}\cdot {\bf E})+ ({\bf \gamma}\cdot
 {\bf B})\right]{\bf B}
\]
\vspace{-8mm}
\begin{equation}
\label{5}
\end{equation}
\vspace{-8mm}
\[
 -({\bf E}\cdot {\bf B}){\bf \beta}- \frac12 \left( {\bf E}^2-{\bf B}^2 \right){\bf \xi},
\]
\[
 {\bf H}=-\frac{\partial {\cal L}}{\partial \textbf{B}}={\bf B}\left[1+({\bf \alpha}\cdot
 {\bf B})- ({\bf \xi}\cdot{\bf E})\right]
 +\left[({\bf \beta}\cdot {\bf E})+ ({\bf \gamma}\cdot
 {\bf B})\right]{\bf E}
\]
\vspace{-8mm}
\begin{equation}
\label{6}
\end{equation}
\vspace{-8mm}
\[
+({\bf E}\cdot {\bf B}){\bf \gamma}- \frac{1}{2}\left({\bf E}^2-
{\bf B}^2\right){\bf \alpha}.
\]
At the case $\theta_i^{(1)}=\theta_i^{(2)}$ (${\bf \alpha}={\bf
\beta}$), (${\bf \xi}={\bf \gamma}$), we arrive at NC
electrodynamics. The second pair of Maxwell equations $\partial_\mu
\widetilde{F}_{\mu \nu }=0$ reads
\begin{equation}
\frac {\partial}{\partial t}{\bf B}+{\bf \nabla}\times {\bf
E}=0,\hspace{0.3in} {\bf \nabla}\cdot{\bf B}=0 . \label{7}
\end{equation}
Here we study the propagation of a linearly polarized laser beam in
the external transverse magnetic field for the case when the
effective Lagrangian of electromagnetic fields is given by Eq.(1).
This consideration generalizes the results of \cite{Guralnic},
\cite{Cai}, \cite{Abe}, \cite{Zahn}, \cite{Mariz}, \cite{Chatillon}
in the case of two independent antisymmetric ``tensor"-parameters
$\theta^{(1)}_{\alpha\beta}$ and $\theta^{(2)}_{\alpha\beta}$.

\section{Energy-Momentum Tensor}

Now, we find the energy-momentum tensor to clear up the direction of
energy propagation. With the help of the standard procedure
\cite{Landau}, we obtain the gauge-invariant canonical
energy-momentum tensor of electromagnetic fields
\[
T_{\mu\nu}=-F_{\mu\alpha}F_{\nu\alpha}\left(1-\frac12\theta^{(1)}_{\gamma\beta}
 F_{\gamma\beta}\right)+\frac14
 \theta^{(1)}_{\mu\alpha}F_{\nu\alpha}F^2_{\rho\beta}
\]
\vspace{-8mm}
\begin{equation}
\label{8}
\end{equation}
\vspace{-8mm}
\[
-\theta^{(2)}_{\mu\beta}F_{\gamma\nu}
F_{\rho\beta}F_{\gamma\rho}-\left(
F_{\mu\alpha}F_{\nu\gamma}+F_{\nu\alpha}F_{\mu\gamma}\right)
\theta^{(2)}_{\alpha\beta}F_{\gamma\beta}-\delta_{\mu\nu}{\cal L} .
\]
At $\theta^{(1)}_{\mu\beta}=\theta^{(2)}_{\mu\beta}$, the canonical
tensor (8) converts into one for NC electrodynamics, obtained in
\cite{Kruglov1}. Tensor (8) is non-symmetric, but is conserved,
$\partial_\mu T_{\mu\nu}=0$. From Eq.(8), we find the energy density
${\cal E}$, and the Poynting vector $\textbf{P}$:
\[
{\cal E}=T_{44}=\frac {{\bf E}^2 +{\bf B}^2}{2}\left[1+({\bf
\alpha}\cdot {\bf B})\right]-\left({\bf \xi}\cdot{\bf
E}\right)\textbf{E}^2 -\left({\bf E}\cdot{\bf B}\right) \left({\bf
\beta}\cdot{\bf E}\right) ,
\]
\vspace{-8mm}
\begin{equation}
\label{9}
\end{equation}
\vspace{-8mm}
\[
T_{m4}=-iP_m ,~~~~{\bf P}={\bf E}\times{\bf H}.
\]
so that the four-vector of the energy-momentum is $P_\mu=({\bf
P},i{\cal E})$, and the continuity equation $\partial_\mu P_\mu =0$
is valid. With the help of Eq.(6), the Poynting vector can also be
written as
\[
{\bf P}=\left[1+({\bf \alpha}\cdot{\bf B})-\left({\bf \xi}\cdot{\bf
E}\right)\right]({\bf E}\times{\bf B})
\]
\vspace{-8mm}
\begin{equation}
\label{10}
\end{equation}
\vspace{-8mm}
\[
+\frac12 \left({\bf B}^2-{\bf E}^2\right)({\bf E}\times {\bf
\alpha})+(\textbf{E}\cdot\textbf{B})(\textbf{E}\times {\bf \gamma}).
\]
It follows from Eq.(10) that the direction of the Poynting vector
(and the energy propagation) is different from the direction of the
wave vector $\textbf{k}$ or $({\bf E}\times{\bf B})$. So, the vacuum
in the model considered, behaves like an anisotropic medium (see
\cite{Mariz} for a particular case
$\theta^{(1)}_{\alpha\beta}=\theta^{(2)}_{\alpha\beta}$).

The symmetric energy-momentum tensor can be obtained by varying the
action, corresponding to the Lagrangian (1), on the metric tensor
$g^{\mu\nu}$ \cite{Landau}. After calculations, we arrive at the
symmetric energy-momentum tensor:
\begin{equation}
T^{sym}_{\mu\nu}=T_{\mu\nu}+\frac14 \theta^{(1)}_{\nu\alpha}
 F_{\mu\alpha}F^2_{\rho\beta}
-\theta^{(2)}_{\nu\beta}F_{\gamma\mu} F_{\rho\beta}F_{\gamma\rho} ,
\label{11}
\end{equation}
where the conserved tensor $T_{\mu\nu}$ is given by Eq.(8). As the
action corresponding to the Lagrangian (1) is not a scalar, the
conservation of the symmetrical energy-momentum tensor obtained (11)
is questionable. From Eq.(8), (11), one can obtain non-zero traces
of the canonical and symmetrical energy-momentum tensors. For
classical electrodynamics,
$\theta^{(1)}_{\alpha\beta}=\theta^{(2)}_{\alpha\beta}=0$, and
therefore the trace of the canonical energy-momentum tensor
vanishes. It should be noted that the modified energy-momentum
tensor leads to changing the curvature of space-time. This may have
an influence on the inflation of the universe.

\section{Vacuum Birefringence}

Now we consider the plane electromagnetic wave
($\textbf{e},\textbf{b}$) propagating in $z$-direction and
perpendicular to the external constant and uniform magnetic field
$\overline{\textbf{B}}=(\overline{B},0,0)$. Then
$\textbf{E}=\textbf{e}$,
$\textbf{B}=\textbf{b}+\overline{\textbf{B}}$. The rotation of the
magnetic field, in the PVLAS experiment, does not effect the vacuum
birefringence within QED calculations \cite{Adler}, \cite{Biswas},
and therefore, we consider the stationary and uniform external
magnetic field. After linearizing Eq.(5),(6) around the background
magnetic induction field $\overline{\textbf{B}}$, one obtains the
linearized equations:
\begin{equation}
d_i=\varepsilon_{ij}e_j+\rho_{ij}b_j
,~~~~h_i=(\mu^{-1})_{ij}b_j+\sigma_{ij}e_j\label{12}
\end{equation}
where
\[
\varepsilon_{ij}=\left[1+({\bf \alpha}\cdot
\overline{\textbf{B}})\right]\delta_{ij} -
\beta_i\overline{B}_j-\beta_j\overline{B}_i,~~\rho_{ij}=\xi_i\overline{B}_j-
\delta_{ij}({\bf \gamma}\cdot
\overline{\textbf{B}})-\overline{B}_i\gamma_j,
\]
\vspace{-7mm}
\begin{equation}  \label{13}
\end{equation}
\vspace{-7mm}
\[
\mu^{-1}_{ij}=\left[1+({\bf \alpha}\cdot
\overline{\textbf{B}})\right]\delta_{ij} +
\alpha_i\overline{B}_j+\alpha_j\overline{B}_i,~~~~
\sigma_{ij}=-\overline{B}_i \xi_j+ \delta_{ij}({\bf \gamma}\cdot
\overline{\textbf{B}})+\gamma_i \overline{B}_j.
\]
From Eq.(12) and Maxwell equations
\[
k_i d_i=k_i b_i=0,
\]
\vspace{-7mm}
\begin{equation}  \label{14}
\end{equation}
\vspace{-7mm}
\[
\textbf{k}\times\textbf{e}=\omega\textbf{b},~~~~\textbf{k}\times\textbf{h}=-\omega\textbf{d},
\]
we find the equation for the electric field $\textbf{e}$:
\[
\biggl[\textbf{k}^2\left(\mu^{-1}\right)_{bi}+k_a\left(\mu^{-1}\right)_{al}k_l\delta_{ib}
-\textbf{k}^2\left(\mu^{-1}\right)_{pp}\delta_{ib}-k_l\left(\mu^{-1}\right)_{bl}k_i
\]
\vspace{-7mm}
\begin{equation}  \label{15}
\end{equation}
\vspace{-7mm}
\[
+ \omega^2\varepsilon_{ib}+\omega\varepsilon_{ijk}k_j\sigma_{kb}
+\omega\rho_{ij}\varepsilon_{jmb}k_m \biggr]e_b=0 ,
\]
where $\varepsilon_{ijk}$ is the antisymmetric tensor
($\varepsilon_{123}=1$). The homogeneous Eq.(15) possesses
non-trivial solutions when the determinant of the matrix equals
zero. To simplify the problem, we consider the case
$\theta^{(1)}_{4\beta}=\theta^{(2)}_{4\beta}=0$ (${\bf \xi}={\bf
\gamma}=0$). It should be mentioned that NC field theory preserves
unitarity only for non-zero space-space non-commutativity,
$\theta_{0a}=0$ \cite{Gomis}, \cite{Aharony}. In addition, the
bounds on the time-space components $\theta_{0a}$ are much weaker
($\theta^{-1/2}> {\cal O}
 (10~GeV)$) compared to space-space components ($\theta^{-1/2}>{\cal O}(10~TeV)$)
\cite{Carroll}. Evaluating the determinant for this case (${\bf
\xi}={\bf \gamma}=0$), we obtain the dispersion relation:
\begin{equation}
A^2\left[A+2n^2{(\bf \alpha}\cdot\overline{\textbf{B}})-2({\bf
\beta}\cdot\overline{\textbf{B}})\right]=0, \label{16}
\end{equation}
where
\begin{equation}
A=1+({\bf \alpha}\cdot\overline{\textbf{B}})-n^2\left[1+3({\bf
\alpha}\cdot\overline{\textbf{B}})\right], \label{17}
\end{equation}
and $n=k/\omega$ is the index of refraction. There are two solutions
to Eq.(16):
\[
A=0,~~~~n_\bot^2=1-2({\bf \alpha}\cdot\overline{\textbf{B}})
\]
\begin{equation}
A+2n^2{(\bf \alpha}\cdot\overline{\textbf{B}})-2({\bf
\beta}\cdot\overline{\textbf{B}})=0, \label{18}
\end{equation}
\[
n_\|^2=1-2({\bf \beta}\cdot\overline{\textbf{B}}).
\]
In Eq.(18), we use the expansion in small parameters ${\bf \alpha}
$, ${\bf \beta} $. The $n_\bot^2$, $n^2_\| $ correspond to the cases
when the electric field of the plane wave $\textbf{e}$ is
perpendicular ($\textbf{e}\bot \overline{\textbf{B}}$) and parallel
($\textbf{e} \| \overline{\textbf{B}}$) to the background magnetic
induction field $\overline{\textbf{B}}$. So, the speed of light is
different for two modes. At the case ${\bf \alpha}={\bf \beta}$, we
arrive at the result \cite{Guralnic}, that the speed of light is
shifted equally for both polarizations. Only at the case ${\bf
\alpha}\neq{\bf \beta}$, we have the effect of the induced
ellipticity or birefringence. In the general case ${\bf \xi}\neq 0$,
${\bf \gamma}\neq 0$, parameters ${\bf \xi}$, ${\bf \gamma}$
contribute to indexes of refraction $n_\bot$, $n_\|$. If the angle
between the polarization vector $\textbf{e}$ and the external
magnetic induction field $\overline{\textbf{B}}$ is $\theta$, then
the polarization vector at $z=0$ is
$\textbf{e}|_{z=0}=E_0(\cos\theta, \sin\theta)\exp(-i\omega t)$. The
components of the polarization vector at arbitrary $z$ are given by
\begin{equation}
e_\bot=E_0\sin\theta\exp i\left(k_\bot z-\omega
t\right),~~~~e_\|=E_0\cos\theta\exp i\left(k_\| z-\omega t\right),
\label{19}
\end{equation}
where $k_\bot=n_\bot \omega$, $k_\|=n_\| \omega$. We obtain from
Eq.(19) \cite{Born}
\begin{equation}
\alpha=\theta, ~~~~\delta=\left(k_\bot-k_\|\right)z=\left(\left({\bf
\beta}-{\bf \alpha}\right)\cdot\overline{\textbf{B}}\right)\omega
z,~~\sin2\chi= \left(\sin2\alpha \right)\sin\delta . \label{20}
\end{equation}
One finds from Eq.(20) the induced ellipticity (the ratio of minor
to major axis of the ellipse)
\begin{equation}
\Psi\equiv\tan\chi\simeq\chi\simeq \frac{1}{2}\delta\sin2\theta
=\frac{\left(\left({\bf \beta}-{\bf
\alpha}\right)\cdot\overline{\textbf{B}}
 \right)\pi L}{\lambda}\sin2\theta ,
 \label{21}
\end{equation}
where $\omega=2\pi/\lambda$, $\lambda$ is a wave length. We have
used here the smallness of the $\delta$. As a result, after
propagating the distance $L$, initially linearly polarized light
becomes elliptically polarized. One obtains to first order in the
small parameter $\delta$: $\psi\simeq\theta$ (because the angle of
the rotation of the ellipse $\psi$ is given by
$\tan2\psi=(\tan2\alpha)\cos\delta$). There is no rotation of the
polarization axis of the ellipse.

With the help of the preliminary results of the PVLAS experiment
\cite{Zavattini1}
\[
\Psi=(-3.4\pm0.3)\times10^{-12}\frac{{\mbox r}{\mbox a}{\mbox
d}}{{\mbox p }{\mbox a}{\mbox s}{\mbox s}},~~L=1~{\mbox m},
\]
\vspace{-7mm}
\begin{equation}  \label{22}
\end{equation}
\vspace{-7mm}
\[
\lambda=1064~ {\mbox n}{\mbox
m},~~\theta=\frac{\pi}{4},~~\overline{B}=5.5 ~{\mbox T}
,~~e\overline{B}=3.25\times10^{-10}~ (MeV)^2,
\]
one can find from Eq.(21) the constraint for the parameter
difference:
\begin{equation}
\left({\bf \beta}-{\bf
\alpha}\right)_{\overline{\textbf{B}}}\simeq10^{-9}~(MeV)^{-2}
,\label{23}
\end{equation}
where the subscript means the projection on the direction of
$\overline{\textbf{B}}$. The induced ellipticity of the PVLAS
experiment can be explained within the effective Lagrangian (1).
Possibly the improvement of PVLAS dada will change the estimation
(23) to the lower value.

\section{Conclusion}

We suggest the effective Lagrangian at the cubic order in the
electromagnetic field strength which contains two ``tensors"
$\theta^{(1,2)}_{\mu\nu}$. This is a generalization of NC
electrodynamics. At the limit
$\theta^{(1)}_{\mu\nu}=\theta^{(2)}_{\mu\nu}$, one arrives at
electrodynamics on NC spaces (with the help of the Seiberg-Witten
map). The Lorentz covariance is broken because parameters
$\theta^{(1,2)}_{\mu\nu}$ are not transformed as real tensors.
Lorentz violating structures at the quadratic order in field
strength, leading to birefringence in a vacuum without a magnetic
field, were discussed in \cite{Kostelecky}. The Lorentz violating
operators at quadratic order are constrained by astrophysical data
\cite{Kostelecky1}, \cite{Kostelecky2}, and therefore, we do not
include these structure in the Lagrangian investigated.

The density of the energy and momentum, and the canonical and
symmetric energy-momentum tensors are found. The canonical
energy-momentum tensor is conserved, but the symmetric
energy-momentum tensor, obtained by varying the action on the metric
tensor, is non-conserved. The traces of the canonical and symmetric
energy-momentum tensors do not equal zero, i.e., there is a trace
anomaly at the tree level. This anomaly is related to the violation
of the Lorentz invariance. We show that the propagation of the
electromagnetic wave in the constant magnetic background and the
energy propagation have different directions, i.e. the vacuum is
similar to an anisotropic medium.

It was proven that the model suggested leads to the induced
ellipticity which, at the case ${\bf \alpha}={\bf \beta}$, ${\bf
\gamma}={\bf \xi}=0$, disappears in accordance with the previous
results \cite{Guralnic}, \cite{Cai}, \cite{Abe}, \cite{Zahn},
\cite{Mariz}, \cite{Chatillon}. We have calculated the induced
ellipticity through the parameters ${\bf \alpha} $, ${\bf \beta}$.
This relation allows us to obtain the constraint on parameters
introduced to explain the ellipticity observed in the PVLAS
experiment. For the case  ${\bf \xi}\neq 0$, $\gamma\neq 0$, the
induced ellipticity depends on four ``vector"-parameters ${\bf
\alpha} $, ${\bf \beta}$, ${\bf \xi}$, $\gamma$.

It should be mentioned that the discussed additions to the
Lagrangian can not explain all of the PVLAS observations because
they do not lead to a rotation of the polarization (dichroism). We
leave the discussion of bounds on parameters introduced coming from
astrophysics for further investigations.

\textsc{\textbf{Acknowledgements}}

I would like to thank Prof. V. A. Kostelesk\'{y} for communications.

\textit{Added note.} - After this paper was prepared, I studied the
manuscript E. Zavattini et al. [PVLAS Collaboration],
arXiv:0706.3419 [hep-ex] with new experimental data. New PVLAS
results give limits on magnetically induced rotation and ellipticity
in vacuum and correct Eq.(22) and our estimation (23) to the lower
value.

\end{document}